%% file: 0_main.tex
\documentclass[sigconf]{acmart}
\acmConference[EASE 2026]{The 30th International Conference on Evaluation and Assessment in Software Engineering}{9–12 June, 2026}{Glasgow, Scotland, United Kingdom}

\input{macro}

\AtBeginDocument{%
  }

\begin{document}

\title{Verifier Warnings Do Not Improve Comprehensibility Prediction}

\author{Nadeeshan De Silva}
\affiliation{%
  \institution{William \& Mary}
  \city{Williamsburg}
  \state{Virginia}
  \country{USA}
}
\email{kgdesilva@wm.edu}

\author{Martin Kellogg}
\affiliation{%
  \institution{New Jersey Institute of Technology}
  \city{Newark}
  \state{New Jersey}
  \country{USA}
}
\email{martin.kellogg@njit.edu} 

\author{Oscar Chaparro}
\affiliation{%
  \institution{William \& Mary}
  \city{Williamsburg}
  \state{Virginia}
  \country{USA}
}
\email{oscarch@wm.edu}

\begin{abstract}
  Proponents of software verification suggest that code simplicity is
  linked to the effort to verify code, hypothesizing that formal
  verifiers produce fewer false positive warnings and require
  less manual intervention when analyzing simpler code. A recent
  meta-analysis study found empirical support for this hypothesis: a
  small correlation between the sum of verifier warnings and
  human-derived code comprehensibility metrics. Based on this finding,
  we conjectured that using the sum of verifier tool (verifier) warnings to
  represent program semantic information as an input feature to
  machine learning (ML) models for code comprehensibility
  prediction can enhance their performance, when combined with
  traditional syntactic and developer features.
  
  To test this conjecture, 
 we performed a control-treatment experiment incorporating the
 \textit{verifier warning sum} feature into machine learning
 models from the literature, and conducted a comparative analysis of their performance
 against models trained only on syntactic and developer features.  We
 found no significant difference in the prediction performance of
 models with and without the warnings feature.  Our findings suggest
 that while a correlation exists, the \textit{verifier warning
   sum} offers limited discriminative power:
 combining syntactic and developer features is just as effective for predicting human-judged code comprehensibility.

\end{abstract}

\maketitle

\input{1_intro}

\input{5_related_work}

\input{3_methodology}

\input{4_results}

\input{8_threats_to_validity}

\input{6_conclusions}

\input{7_acknowledgements}

\balance
\bibliographystyle{ACM-Reference-Format}
\bibliography{references}

\end{document}

%% file: macro.tex
\usepackage[utf8]{inputenc}
\usepackage{url}
\usepackage{colortbl}
\usepackage{balance}
\usepackage{xspace}
\usepackage{graphicx}
\usepackage{array} 
\usepackage{verbatim}
\usepackage{rotating}
\usepackage{bold-extra}
\usepackage{multirow}
\usepackage{listings}
\usepackage{boxedminipage}
\usepackage{soul}
\usepackage{enumitem}
\usepackage[normalem]{ulem}
\setlist{nolistsep,leftmargin=.5cm}
\usepackage{booktabs}
\usepackage{tabularx}
\usepackage{svg}
\usepackage{calc}
\usepackage{float}
\usepackage{multirow}
\usepackage[normalem]{ulem}
\useunder{\uline}{\ul}{}
\usepackage[most]{tcolorbox}
\definecolor{MidnightBlue}{HTML}{006895}
\definecolor{BoxesBlue}{HTML}{DEECFF}
\definecolor{BoxesYellow}{HTML}{FFF2CC}
\definecolor{StateGreen}{HTML}{91C788}
\definecolor{StateRed}{HTML}{FF8080}
\definecolor{ArrowGreen}{HTML}{61B15A}
\definecolor{ArrowViolet}{HTML}{BA94D1}
\usepackage{caption}
\usepackage[compact]{titlesec}
\usepackage{microtype} 
\usepackage{amsmath,amsfonts}
\usepackage{algorithmic}
\usepackage{textcomp}
\usepackage{xcolor}
\usepackage{hyperref}
\usepackage{ifthen}
\usepackage{wrapfig}
\usepackage{subcaption}
\usepackage{cleveref}

\usepackage{threeparttable}

\newboolean{showcomments}
\setboolean{showcomments}{true}

\ifthenelse{\boolean{showcomments}}
{\newcommand{\nb}[2]{
		\fbox{\bfseries\sffamily\scriptsize#1}
		{\sf\small$\blacktriangleright$\textit{#2}$\blacktriangleleft$}
	}
}
{\newcommand{\nb}[2]{}
}

\newcommand{\ie}{\textit{i.e.},\xspace}
\newcommand{\eg}{\textit{e.g.},\xspace}

\newcommand{\etal}{\textit{et al.}\xspace}
\newcommand{\aka}{\textit{a.k.a.}\xspace}

\usepackage{tikz}

\definecolor{bug_red}{rgb}{.84,.23,.29}

\definecolor{info-needed-color}{rgb}{1,.8,.12}

\definecolor{lightblue}{rgb}{ .753,  .902,  .961}

\definecolor{lightgreen}{rgb}{.596, 1, .596}
\DeclareRobustCommand{\hlgreen}[1]{{\sethlcolor{lightgreen}\hl{#1}}}

\definecolor{lightyellow}{rgb}{1,0.94902,0.8}

\newcommand{\ABUFIF}{{ABU}\textsubscript{50\%}\xspace}

\newcommand{\BDFIF}{{BD\textsubscript{50\%}}\xspace}
\newcommand{\PBU}{{PBU}\xspace}
\newcommand{\AU}{{AU}\xspace}
\newcommand{\RL}{{RL}\xspace}

\newcommand{\DIFF}{\boldmath{$\Delta_{wF1}$}\xspace}
\newcommand{\DIFFAUC}{\boldmath{$\Delta_{wAUC}$}\xspace}

\newcommand{\Code}{\boldmath{$C$}\xspace}
\newcommand{\CW}{\boldmath{$T$}\xspace}

\newcommand{\dsthree}{\boldmath{{$DS_2$}}\xspace}
\newcommand{\dssix}{\boldmath{{$DS_1$}}\xspace}

%% file: 1_intro.tex
\section{Introduction}
\label{sec:intro}

In order to perform software engineering tasks such as feature development, code review, refactoring, or bug fixing, developers must deeply understand the code they are working with~\cite{Borstler:2023developers, Maalej:TOSEM14, Minelli:ICPC15, Xia:TSE18}. 
Gaps in such understanding can lead to introducing defects, degrading code quality, and ultimately increasing code maintenance effort~\cite{Tao:FSE12, Ko:JVLC2005, Eick:TSE2002}.
While understanding existing code is essential, it is often challenging and time-consuming; prior studies estimate that developers spend more than half of their time trying to understand existing code~\cite{Zuse:IWCP'93, Minelli:ICPC15}.

Researchers have proposed various mechanisms to measure code comprehension effort (\aka code comprehensibility), to control for code that is hard to understand during project evolution.
One mechanism is directly asking humans and collecting human judgments during user studies. However, measuring how well developers understand code is difficult: human judgments are subjective, vary widely, and are expensive to collect~\cite{Wyrich:ACS23,Siegmund:SANER16, Feitelson:ICPC2021, De:2025relative}. %
To address this challenge, researchers introduced objective proxies~\cite{Scalabrino:TSE19,Raymond:TSE10}, such as structural measures (\eg lines of code and comment presence) and complexity metrics (\eg McCabe’s~\cite{McCabe:TSE76} and Halstead’s~\cite{Halstead1977}). 
However, these metrics do not correlate strongly with comprehension measures collected from humans~\cite{Scalabrino:TSE19, Peitek:ICSE21, Feigenspan:ESEM11}.

This mismatch led to researchers proposing machine-learning (ML) approaches that leverage combinations of metrics (\aka features) to approximate human comprehensibility~\cite{Scalabrino:TSE19, Raymond:TSE10, Trockman:MSR18, Lavazza:ESE23}.
For example, the most recent study~\cite{Scalabrino:TSE19} evaluated six ML models trained on 121 syntactic and developer-related features to predict human comprehensibility proxies such as \textit{perceived binary understandability}. A current challenge in the research community is to understand what factors could make these models predict code comprehensibility more accurately. In this work, we contribute to this goal by reporting the results of our investigation on leveraging information produced by formal code verification tools (\emph{verifiers}) to predict human comprehension effort.
\looseness=-1

Our work is motivated by a recent meta-analysis study by Feldman \etal ~\cite{Feldman:FSE23}, which investigated the relationship between formal code verification and human code comprehensibility. Using six existing comprehensibility datasets and four Java verifiers,
which check for code correctness properties (\eg absence of null dereferences),  
they found  evidence that false alarms from such tools may signal when the code is harder to understand for humans. In particular, they showed that such tools produce fewer false positives and require less human intervention on simpler code, reporting a small correlation between the \textit{sum of verifier warnings} from the four verifiers and 20 human comprehensibility proxies.

Intuitively, when code becomes more complex, it makes it much harder for developers to understand it~\cite{Ajami:EMSE19,Scalabrino:TSE19,Peitek:ICSE21,Antinyan:EMSE17,Antinyan:IEEE20}.
Formal verifiers are designed to handle a certain threshold of code complexity before issuing warnings. From an algorithmic perspective, performing any task requires a baseline level of inherent logic, known as essential complexity~\cite{brooks1987no}. However, if the code can be simplified to eliminate a verification warning without altering its underlying functionality, it indicates that the original implementation contained unnecessary, or accidental, complexity (\ie it was unnecessarily complex).
Complex code tends to obscure logical errors, safety issues, and subtle run-time risks, which formal verifiers aim to detect.
As complexity increases, these tools tend to issue more (false positive) warnings and/or require more intervention from human operators (\eg type annotations or loop invariants), reflecting the higher potential for subtle defects and verification challenges.

Based on Feldman \etal's findings~\cite{Feldman:FSE23}, in this work, we conjecture that verifier warnings could be leveraged as a semantic signal for automatically measuring code comprehensibility via machine learning (ML).
To test this conjecture, we designed a control–treatment experiment~\cite{Sjoberg:TSE2005} to investigate whether incorporating the \textit{verifier warning sum} as a program semantic feature enhances the predictive performance of ML models used in prior work~\cite{Scalabrino:TSE19}. 
 
 Specifically,  in the \textbf{Control} setting, we trained six ML models (\eg Random Forests and Support Vector Machines) using only the syntactic and developer features extracted from two prior datasets \cite{Scalabrino:TSE19,Raymond:TSE10}. These  are the largest comprehensibility datasets available, used by Feldman \etal~\cite{Feldman:FSE23}: they contain 150 Java code methods and corresponding 13,860 human comprehensibility measures for five proxies. Using a nested cross-validation approach~\cite{cross_validation}, we evaluated the accuracy of the models in predicting these proxies. In the \textbf{Treatment} setting, we introduced the \textit{verifier warning sum} as a new semantic feature and retrained the same models to predict the same comprehensibility proxies. This semantic feature was obtained by executing four state-of-the-art Java code verifiers (\eg OpenJML~\cite{openjml} and the Checker Framework~\cite{Papi:ISSTA2008}), also used in the prior study by Feldman \etal 

Our comparative analysis of model predictive performance between control and treatment settings (\cref{sec:code_comprehensibility_results}) found that adding the verifier warning sum does not significantly improve the predictive performance of code comprehensibility models.
We further analyzed the impact of individual verifier-specific warnings (\cref{sec:tool_specific_analysis}) and observed that none of the model-metric-verifier combinations led to consistent, statistically significant improvements in model performance; this aligns with our observations regarding the verifier warning sum.
Our findings highlight several promising directions for future work (see \cref{sec:conclusions}), including the need to explore richer semantic representations and other factors that shape code comprehensibility to build more reliable prediction models.

We provide an online replication package~\cite{repl_pack} that contains our dataset,
code, experimental infrastructure and results to facilitate verification and
replication of our study.

%% file: 5_related_work.tex
\section{Background and Related Work}
\label{sec:related_work}

This section reviews prior work on code comprehensibility proxies and predictive models of human understandability, and examines the relationship between formal verification and code comprehension, providing background for our work.

\textbf{Empirical studies of code comprehensibility.}
Researchers have introduced various proxies to measure the difficulty to understand code, since directly measuring how humans understand code is difficult.  
The most commonly used proxies are related to \textit{code correctness} (\eg number of questions about code answered correctly)~\cite{Scalabrino:TSE19}, \textit{subjective ratings} (\eg Likert scale of how well developers understand the code)~\cite{Raymond:TSE10}, \textit{time} (\eg time to read/understand code), and \textit{physiological} metrics. 
These physiological measures include eye-tracking data (\eg gaze duration, fixation count)~\cite{Karas:TSE2024, Abbad-Andaloussi:ICPC22, Binkley:EMSE13, Fritz:ICSE14, Peitek:ICPC20, Park:ESE24}, biometric sensory data~\cite{Fritz:ICSE14,Fucci:ICPC19,Yeh:FIE17}, fMRI scanner measurements~\cite{Peitek:ICSE21, Peitek:TSE18, Siegmund:ICSE14}, and heart rate sensor data~\cite{soga2025your}.
\looseness=-1

Prior studies have found that traditional code metrics, such as McCabe's cyclomatic complexity~\cite{McCabe:TSE76}, Halstead volume~\cite{Halstead1977}, coupling/cohesion~\cite{stevens1974structured}, and maintainability index~\cite{Oman:1992}, poorly correlate with these proxies~\cite{Ajami:EMSE19, Feigenspan:ESEM11, Jbara:EMSE17, Kaner2004, Scalabrino:TSE19, Peitek:ICSE21}. 
For example, Scalabrino \etal~\cite{Scalabrino:TSE19} found only a small correlation between code- and developer-related metrics collected from developers and students using open-source projects. 
Furthermore, the findings of Peitek \etal~\cite{Peitek:ICSE21} suggest that code complexity metrics such as cyclomatic complexity are ineffective for measuring code comprehensibility.

\textbf{Predictive models of understandability.} The low correlation between traditional metrics and comprehensibility proxies has motivated the development of predictive models of comprehensibility, which directly predict human-judged proxies based on code properties (\eg lines of code or cyclomatic complexity) and developer information (\ie programming experience).  These models aim to predict how difficult it is for a particular developer to understand a given code snippet. 

Numerous studies have proposed machine learning models for predicting comprehensibility~\cite{Scalabrino:TSE19, Trockman:MSR18, Lavazza:ESE23, Raymond:TSE10, Posnett:SEN21, dorn2012general, Scalabrino:ICPC16, Scalabrino:JSEP18}, which is typically formulated as a classification problem (\eg predict whether the code is ``easy'' or ``hard'' to understand).  Some studies have developed regression models~\cite{Scalabrino:TSE19,Lavazza:ESE23,Trockman:MSR18, lavazza2023empirical} (\eg predict comprehension time), but their accuracy is typically lower due to the inherent noise of continuous proxies, stemming primarily from hard to control factors during user studies (\eg different pace of participants when reading and understanding the code). 
Model inputs in these works include structural features (\eg loops, operators, blank lines)~\cite{Raymond:TSE10, Posnett:SEN21}, aggregated features such as visual matrix representations of code tokens and alignment-based metrics~\cite{dorn2012general}, and lexical features including comment readability and textual coherence introduced in more recent studies~\cite{Scalabrino:ICPC16, Scalabrino:JSEP18}.

Among the prior studies, Scalabrino \etal~\cite{Scalabrino:TSE19} present the current state of the art in the area, as they developed several ML models (\eg Random Forests and Support Vector Machines) that outperform earlier approaches~\cite{Trockman:MSR18}.
In this work, we build upon Scalabrino \etal's methodology to develop and evaluate ML models of comprehensibility, to test whether semantic information from code verifiers improves model accuracy.

\begin{figure*}[t]
	\centering
	\includegraphics[width=\textwidth]{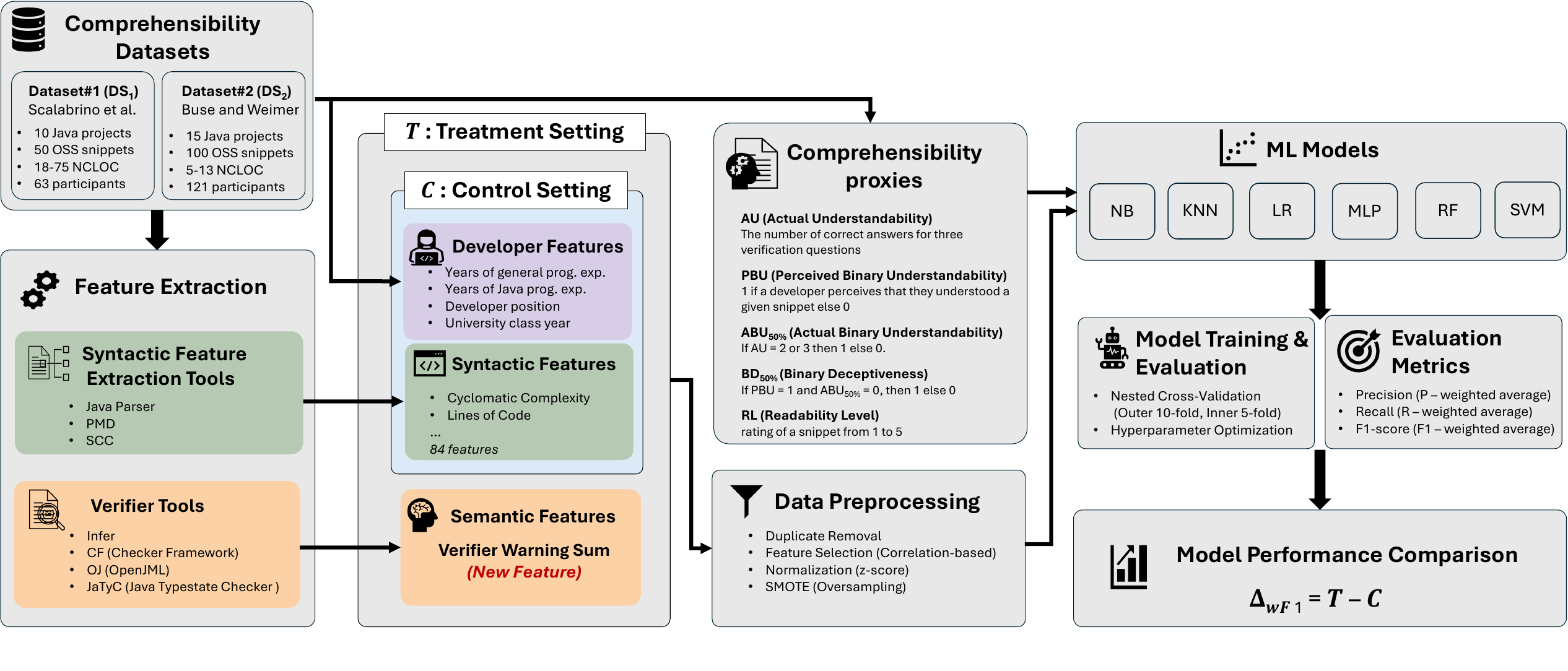}
	\caption{Overview of our methodology for evaluating the impact of verifier warnings on code comprehensibility prediction. We compare two model variants: a \textit{control} model trained on syntactic and developer features, and a \textit{treatment} model trained on the same features plus the verifier warning sum as a semantic feature. Both models are evaluated using nested cross-validation.}
	\label{fig:methodology_overview}
\end{figure*}

With the emergence of LLMs for code generation, recent research has examined the readability of AI-generated code~\cite{Sergeyuk:ICPC24, Patel:24BigData, Dantas:2023}.
Dantas \etal~\cite{Dantas:2023} performed an analysis comparing code generated by ChatGPT with human-written code from Stack Overflow. They evaluated the code snippets using the SonarQube tool to identify and analyze potential code readability issues. 
Sergeyuk \etal~\cite{Sergeyuk:ICPC24} performed a comparison against existing code  readability models (\eg~\cite{Scalabrino:TSE19,Posnett:MSR2011, Mi:JSS22}) on AI-generated Java snippets and evaluated the same snippets with a human study, finding that existing models poorly correlate  with human judgments---making those models unsuitable as proxies for human evaluation of AI-generated code.
Patel \etal ~\cite{Patel:24BigData} compare AI-generated (Microsoft Copilot) and human-written Java solutions to LeetCode problems using static analysis tools (Understand, SpotBugs, PMD---none of which are verifiers), finding that AI code tends to be longer but has lower bug density than human code.
None of these latest studies introduces a predictive model of code comprehensibility, and hence we decided to work with models introduced by Scalabrino \etal~\cite{Scalabrino:TSE19}. However, these studies emphasize the importance of code comprehensibility prediction models and the need to improve them to better estimate human comprehension of code, thereby supporting practical software maintenance tasks more effectively.

\textbf{Formal code verification and comprehensibility.}
Recent research by Feldman \etal~\cite{Feldman:FSE23} investigates the relationship between code verifiability and human understandability. 
Our work is directly motivated by their work, as they concluded that \textit{``our work has implications for the users and 
designers of verification tools and for future attempts to automatically 
measure code comprehensibility''}. 
They conducted a meta-analysis study to measure the correlation between human-judged code comprehensibility and the difficulty of formally verifying code (\ie verifiability). Using six comprehensibility datasets from prior studies, spanning 18,005 human-collected comprehensibility measurements for  211 Java methods, the study compared the number of ``false positive'' warnings generated by four state-of-the-art static verifiers against 20 human-based comprehensibility proxies of different categories (proxies related to correctness, time, and physiological measurements). 
The study established a measurable but small correlation (Pearson's $r=0.23$) between the two domains, implying that code which is easier/difficult for formal tools to verify is easier/difficult for human developers to comprehend. 
Verifiers inherently must reason about program semantics to prove safety properties; therefore, when they fail and issue ``false positive'' warnings due to convoluted logic or missing specifications, 
these warnings can act as a proxy for semantic code complexity.
The study speculated that verification warnings might capture a semantic dimension of cognitive burden that syntactic metrics miss, suggesting that verifiability could be a valuable input for automated code understandability models.

A correlation between code verifiability and human comprehensibility may mean that both humans and formal verifiers struggle with the same underlying factors that cause code complexity.
Syntactic metrics often assign different scores to code snippets that look different in structure but implement the same algorithm (\eg the same sorting algorithm can be implemented with imperative loops or in a functional style), but verifier warnings could capture the ``accidental complexity'' and unstated assumptions that confuse human developers.

\input{latex_tables/ac_class_distribution}

For example, consider accessing an array using a potentially out-of-bounds index in a Java-like language. A simple bounds check might use an if statement to ensure the index falls within the array's valid range. A more complex variant with the same semantics might access the array inside a try block, relying on a catch statement to intercept the resulting exception if the index proves invalid. This more convoluted approach introduces  accidental complexity and might fail verification: the out-of-bounds access actually occurs, but it is intercepted before crashing the program.
To avoid issuing a false positive warning, a verifier would need to accurately model this exceptional control flow (and, in practice, few verifiers do). Alternatively, a verifier might flag code that relies on unstated assumptions. For instance, accessing an array without a prior bounds check inherently assumes the index has already been validated elsewhere. A verifier might warn that this operation is unsafe unless a developer provides an explicit specification guaranteeing the index's validity for that specific array. With this specification in place, the verifier can approve the array access, but it must then ensure that every integer value assigned to the index variable strictly adheres to that valid range. 
Thus, a warning triggered by a missing specification often highlights underlying code complexity: a human reader would similarly need to reason through the program's logic to determine exactly why that specific array access is safe.

We define the \textit{verifier warning sum} as the summation of false positive warnings generated by the four verifiers used in Feldman \etal's prior study~\cite{Feldman:FSE23} during the analysis of a code snippet.

%% file: latex_tables/ac_class_distribution.tex
\begin{table*}[t]
\centering
\caption{Datasets and Code comprehensibility metrics. NCLOC = Non-Comment/blank Lines of Code}
\label{tab:absolute_proxies}
\resizebox{0.85\textwidth}{!}{
\begin{tabular}{p{3.2cm}|l|l|lcrr}
\toprule
\multicolumn{1}{c|}{\textbf{Dataset}} & \multicolumn{1}{c|}{\textbf{Metric}} & \multicolumn{1}{c|}{\textbf{Definition}}                                                                                                                 & \multicolumn{4}{c}{\textbf{Class Distribution}}                                                                                                                                                                                                                                                   \\ 
\hline
\multirow{4}{=}{\dssix\\
(Scalabrino \etal~\cite{Scalabrino:TSE19})\\
50 OSS snippets of 18 - 75 NCLOC; \\ 50 students and 13 developers
 }
                                & \AU (Actual Understandability)                  & \begin{tabular}[c]{@{}l@{}}The number of correct answers for three verification  \\ questions about the code. Possible values are 0,1,2,3.\\ (higher number implies higher understandability) \end{tabular}    & \begin{tabular}[c]{@{}l@{}}0\\ 1\\ 2\\ 3\end{tabular}      & \begin{tabular}[c]{@{}c@{}}-\\ -\\ -\\ -\end{tabular}     & \begin{tabular}[c]{@{}r@{}}153\\ 72\\ 138\\ 77\end{tabular}             & \begin{tabular}[c]{@{}r@{}}(34.7\%)\\ (16.4\%)\\ (31.4\%)\\ (17.5\%)\end{tabular}            \\ 
\cline{2-7}
                                & \PBU (Perceived Binary Understandability)       & \begin{tabular}[c]{@{}l@{}}1 if a developer perceives that they understood \\ a given snippet; 0 otherwise. \\ (higher number implies higher understandability) \end{tabular}                                     & \begin{tabular}[c]{@{}l@{}}0 \\ 1\end{tabular}              & \begin{tabular}[c]{@{}c@{}}-\\ -\end{tabular}             & \begin{tabular}[c]{@{}r@{}}136\\ 304\end{tabular}                       & \begin{tabular}[c]{@{}r@{}}(30.9\%)\\ (69.1\%)\end{tabular}                                  \\ 
\cline{2-7}
                                & \ABUFIF (Actual Binary Understandability)       & \begin{tabular}[c]{@{}l@{}}If \AU $=$ 2 or 3 then 1 else 0. \\ (higher number implies higher understandability) \end{tabular}                                                                                                                                   & \begin{tabular}[c]{@{}l@{}}0\\ 1\end{tabular}               & \begin{tabular}[c]{@{}c@{}}-\\ -\end{tabular}             & \begin{tabular}[c]{@{}r@{}}225\\ 215\end{tabular}                       & \begin{tabular}[c]{@{}r@{}}(51.1\%)\\ (48.9\%)\end{tabular}                                  \\ 
\cline{2-7}
                                & \BDFIF (Binary Deceptiveness)                   & \begin{tabular}[c]{@{}l@{}}If \PBU = 1 and \ABUFIF = 0, then 1 else 0 \\ (higher number implies lower understandability) \end{tabular}                                                                         & \begin{tabular}[c]{@{}l@{}}0 \\ 1\end{tabular}              & \begin{tabular}[c]{@{}c@{}}-\\ -\end{tabular}             & \begin{tabular}[c]{@{}r@{}}351\\ 89\end{tabular}                        & \begin{tabular}[c]{@{}r@{}}(79.8\%)\\ (20.2\%)\end{tabular}                                  \\ 
\hline
\dsthree 
(Buse and Weimer ~\cite{Raymond:TSE10})
100 OSS snippets of 5 - 13 NCLOC; 121 students                        
                                & \RL (Readability Level)                         & \begin{tabular}[c]{@{}l@{}}Readability rating of a snippet from 1 to 5 \\ (higher value implies higher readability)\end{tabular}                                                                               & \begin{tabular}[c]{@{}l@{}}1\\ 2\\ 3\\ 4\\ 5\end{tabular} & \begin{tabular}[c]{@{}c@{}}-\\ -\\ -\\ -\\ -\end{tabular} & \begin{tabular}[c]{@{}r@{}}889\\ 2481\\ 3240\\ 3290\\ 2200\end{tabular} & \begin{tabular}[c]{@{}r@{}}(7.3\%)\\ (20.5\%)\\ (26.8\%)\\ (27.2\%)\\ (18.2\%)\end{tabular}  \\
\bottomrule
\end{tabular}
}
\end{table*}

%% file: 3_methodology.tex
\section{Methodology}
\label{sec:code_comprehensibility_prediction}

This study focuses on ML-based classification models of human-judged code comprehensibility, aiming to
assess their effectiveness when \textit{verifier warning sum} is included as a complementary feature. With this in mind, we define this research question:\looseness=-1

\vspace{0.1cm}
\noindent \textbf{RQ:} \label{rq:verifier_warnings}{\textit{How much more effective are ML models at predicting code comprehensibility when the verifier warning sum is included?}}
\vspace{0.1cm}

To answer this question, we designed the \emph{control–treatment} experiment~\cite{Sjoberg:TSE2005} depicted in \cref{fig:methodology_overview}. 
In the \emph{control setting}, the models are trained using only syntactic and developer features, as in prior work~\cite{Scalabrino:TSE19}. 
In the \emph{treatment setting}, the models are trained on the same datasets, feature extraction methods, model architectures, evaluation protocols, and the same feature set augmented with the newly introduced semantic feature: \textit{verifier warning sum}. This setup enables a controlled comparison between the two model variants and allows us to isolate and evaluate the added contribution of verifier warnings to predictive performance. 

\input{latex_tables/code_features}

\subsection{Comprehensibility Data Sources}
\label{subsec:ac_data_sources}

We reused the two largest Java code comprehensibility datasets from prior studies~\cite{Scalabrino:TSE19, Raymond:TSE10} (see \cref{tab:absolute_proxies}), both of which trained ML models on hand-crafted features for code comprehensibility prediction:

\begin{itemize}
  \item \textbf{Dataset \#1 (\dssix)}, from Scalabrino \etal~\cite{Scalabrino:TSE19}, contains 50 Java Open Source (OSS) methods evaluated by 50 CS students (with intermediate to high programming experience) and 13 professional developers. Participants judged each method’s understandability, answered three comprehension questions, and had their task-completion time recorded. OSS methods were extracted from 10 open source Java projects, including OpenCMS, Jenkins, Spring Framework, Weka, and K-9 Mail.
  \item \textbf{Dataset \#2 (\dsthree)}, from Buse and Weimer~\cite{Raymond:TSE10}, consists of simplified snippets from 100 Java OSS methods. A total of 121 CS students rated snippet readability on a 5-point Likert scale. These snippets were purposely stripped (size of the snippets were in the range of 5-13 NCLOC) of contextual and algorithmic complexity to mitigate confounding factors and isolate low-level readability attributes. These snippets were taken from 15 OSS Java projects, including jUnit, jEdit, SoapUI, Hibernate, Azureus/Vuze, and JasperReports.
\end{itemize}
These two datasets were also used in Feldman \etal's study~\cite{Feldman:FSE23} and were the largest contributors to their meta-analysis results. 
Given their dominant influence on the findings and their substantially larger size compared to the other datasets considered in such a study~\cite{Feldman:FSE23}, which were relatively small, we selected them for this work to ensure robustness and comparability of results.

All comprehensibility measurements were collected for each individual. \dssix had 6 snippets evaluated by 8 programmers (\aka developers)  and 44 snippets by 9 individuals. \dsthree had 121 programmers evaluate each of the 100 snippets. 
In total \dssix includes 440 human measurements for each of the four code comprehensibility metrics, while \dsthree includes 12,100 measurements for the single metric. These measurements were collected with the code comprehensibility metrics described in \cref{subsec:ac_proxies}.

\subsection{Code Comprehensibility Metrics}
\label{subsec:ac_proxies}
We used the code comprehensibility metrics (\aka proxies) introduced in the above two prior studies as prediction targets for our models. These metrics capture different aspects of the participants' code comprehension and fall into three categories: subjective \textit{ratings} (\eg perceived understandability), the \textit{correctness} of a developer’s understanding of program behavior (\eg based on participants' answers to verification questions), and the \textit{time} required to read or understand the code.

We used the four \textit{correctness} metrics  from \dssix~\cite{Scalabrino:TSE19} and the single metric from \dsthree~\cite{Raymond:TSE10}, all defined in \cref{tab:absolute_proxies}: 
\begin{itemize}
	\item Actual Understandability (\textbf{\AU}) reflects the participant's understanding of the code's behavior based on how correct are their answers to three verification questions posed by the researchers.
	
	\item Perceived Binary Understandability (\textbf{\PBU}) captures subjective judgment of comprehensibility: whether a programmer considered they understood a code snippet.
	
	\item Actual Binary Understandability (\textbf{\ABUFIF}), based on \AU, provides a stricter measure of comprehensibility as it considers a snippet easy to understand only when the programmer correctly answers at least two of three verification questions.
	
	\item  Binary Deceptiveness (\textbf{\BDFIF}), based on \PBU and \ABUFIF, highlights potentially misleading or deceptive code as it considers a snippet hard to understand when the programmer perceived they understood the code (\PBU= 1) but answer only one or none of the three verifications correctly (\ABUFIF= 0).
	
	\item Readability Level (\textbf{RL}) reflects participants' subjective assessment of how easy the code is to read and understand based on a 5-point Likert scale.
\end{itemize}
We excluded the two \textit{time} metrics from \dssix as we focused on  classification  rather than regression of continuous time values. Also, these metrics have high variability (\eg  the TNPU metric ranges from 3 to 1,649 seconds), influenced by individual differences such as cognitive speed, code familiarity, and task complexity, making them less reliable for modeling purposes. 
\looseness=-1

Table \ref{tab:absolute_proxies} summarizes the metrics definitions and class distributions.
All metrics yield discrete comprehensibility scores. In total, \dsthree\ includes 12{,}100 \RL measurements, while \dssix\ includes 440 measurements for each of the other four metrics.

Although the datasets may be relatively small, they are well suited for this study because they contain carefully curated code samples paired with controlled and validated human comprehension measurements. 
Moreover, our experimental design, including data preprocessing steps and nested cross-validation, described later in this section, help mitigate potential biases and increase the likelihood that meaningful patterns can be learned from the data.
Prior studies~\cite{Scalabrino:TSE19,Raymond:TSE10} have successfully trained predictive models and conducted correlation analyses~\cite{Scalabrino:TSE19,Feldman:FSE23} on the same datasets, indicating that they contain sufficient signal for modeling code comprehensibility. 
While the dataset size may limit generalizability, it is adequate for the comparative analysis between control and treatment settings, which is the primary goal of this study.

\subsection{Input Features for Models}
\label{subsec:ac_code_features}

Both prior studies~\cite{Scalabrino:TSE19,Raymond:TSE10} leveraged features capturing various aspects of code and developer characteristics to train models for predicting code comprehensibility. 
In this study, we reused the same set of features from these works and additionally introduce a new feature: \textit{verifier warning sum}, used in the prior meta-analysis study~\cite{Feldman:FSE23}. 
These features can be categorized into three groups: \emph{syntactic}, \emph{developer}, and  \emph{semantic features}. 

\subsubsection{Syntactic Features.} 
We began with the 115 features defined by Scalabrino \etal~\cite{Scalabrino:TSE19}. Upon reviewing their data, definitions, and implementations, we excluded 38 features that were ambiguously defined (\eg unclear term ``word'' in ``\# of words''), inapplicable to the available snippets (\eg ``\# of aligned blocks'' for constructors, of which none were present), or missing for more than 30\% of snippets (also excluded in the prior study~\cite{Scalabrino:TSE19}). 
We then added seven complementary features to ensure consistency (\eg adding total counts where only normalized counts existed). 

In total, we used 84 code features, including cyclomatic complexity,  lines of code, and number of loops, parameters, Java keywords, and comments. \Cref{tab:code_features} shows examples of the features across five categories: \emph{complexity}, \emph{size}, \emph{lexicon}, \emph{format}, and \emph{documentation}. 
\textit{Complexity} features include traditional metrics like loop count
and cyclomatic complexity, while \textit{size} features account for parameters, statements, and lines of code. \textit{Format} features capture stylistic elements such as blank lines and parentheses, whereas \textit{lexicon}
features capture vocabulary, including identifiers and keywords. Finally, \textit{documentation} features account for comments and their readability (\eg via the Flesch reading-ease test~\cite{flesch1979write}). See our replication package~\cite{repl_pack} for the full list of features and their definitions.

\subsubsection{Developer Features.}
We reused all developer-related features from prior work~\cite{Scalabrino:TSE19,Raymond:TSE10}, which capture aspects of a programmer’s background and experience. In \dssix, \textit{years of general programming experience} and \textit{years of Java experience} are modeled as discrete variables, each taking integer values from 1 to 10 to represent increasing levels of experience. The dataset also includes \textit{participant role}, an ordinal variable with four possible values: bachelor student, master student, Ph.D. student, and professional developer.
\dsthree\ includes a single feature, \textit{university class year/academic level}, which represents the participant’s college year. This feature is also ordinal, with four levels: first year, second year, third/fourth year, and graduate level. The original data does not distinguish third- and fourth-year participants~\cite{Raymond:TSE10}.

\subsubsection{Semantic Feature.}
\label{subsec:semantic_features}
We consider \textit{verifier warning sum} as a \emph{semantic feature}. Formal verifiers assess the logical correctness of code and its adherence to semantic rules, rather than merely its syntactic form. 
The warnings they generate indicate potential violations of correctness properties, such as null dereferences, dead code, or run-time errors. 
By aggregating these warnings as counts, we obtain a quantitative measure that captures underlying program semantics, reflecting how the code behaves or could behave.

We used the replication package from previous meta-analysis study~\cite{Feldman:FSE23} to extract the verifier warning sum from \dssix and \dsthree.
We utilized the same four state-of-the-art Java formal verifiers and same released versions to analyze the Java code snippets from the two datasets:
\begin{itemize}
  
 \item The Checker Framework (\textbf{CF})~\cite{Papi:ISSTA2008} is a platform for extending Java's type system to verify properties that the compiler does not reason about by default. By integrating custom type qualifiers into the Java compiler, it enables developers to enforce properties like nullability or regular expression validity at compile time. We re-used the same nine typecheckers used by Feldman \etal~\cite{Feldman:FSE23}: checkers for nullness, interning, object construction, resource leaks, array bounds, signature strings,  format strings, regular expressions and optionals. 
 
\item  OpenJML (\textbf{OJ})~\cite{openjml}  is a Java verifier that uses the Java Modeling Language (JML)~\cite{leavens1998jml} to check whether  program’s implementation satisfies its formal specifications. Developers use JML to specify the intended behavior of classes and methods, including preconditions, postconditions, and class invariants. The tool translates the Java code together with its JML specifications into logical statements called verification conditions (VCs). SMT solvers then attempt to mathematically prove that these VCs are valid.
 OpenJML verifies the absence of common programming errors, including out-of-bounds array accesses, null pointer dereferences, and integer overflows and underflows, even without any programmer-written specifications. 
 
 \item The Java Typestate Checker (\textbf{JaTyC})~\cite{jatyc}  implements a formal typestate analysis~\cite{Strom:TSE2012} that extends Java's static type system to monitor the lifecycle states of objects. Unlike standard type systems, JaTyC tracks state transitions (\eg the sequence of closed, open, and closed for a File object) to ensure protocol adherence. This verifier identifies potential null dereferences, protocol violations in object lifecycles (\eg sockets or files), and unauthorized memory accesses. 

\item \textbf{Infer}~\cite{infer} is a bug-finding tool developed by and deployed within Meta, written in OCaml~\cite{ocaml}, that implements separation logic~\cite{ohearn2001local} and bi-abduction~\cite{calcagno2009compositional} to detect common defects such as null pointer dereferences, data races, and resource leaks. Separation logic facilitates scalable, modular reasoning of state mutations, while bi-abduction automates this inference process. Though Infer itself is not \emph{technically} a verifier---its design intentionally omits warnings that are heuristically judged to be likely false positives---we include it because its reasoning is based on a sound core, unlike truly-heuristic bug-finders like SpotBugs~\cite{SpotBugs}.

\end{itemize}

For each snippet, we recorded the total number of warnings issued by each tool and totaled them across all tools; we refer to this aggregated value as the \textit{verifier warning sum}. Feldman \etal ~\cite{Feldman:FSE23} checked that these warnings are really false positives: they indicate \emph{possible} defects detected by the tools that cannot actually occur if the code is executed.
To ensure consistency, we re-ran the analysis using the data provided in their replication package and identified several discrepancies. 
A major portion of these discrepancies stemmed from warnings generated in artificial stubs rather than in the actual methods under analysis. 
By re-executing the analysis pipeline end to end, we ensured accuracy and replicability instead of relying solely on the published results in the prior study~\cite{Feldman:FSE23}.
We documented these discrepancies in this study's replication package~\cite{repl_pack}.
\looseness=-1

\subsection{Machine Learning Models}
\label{subsec:models}

We used the same six ML models as in prior work~\cite{Scalabrino:TSE19}:
\begin{itemize}
  \item Naïve Bayes (\textbf{NB})~\cite{naive_bayes} predicts the most probable class using Bayes’ theorem under a feature-independence assumption 
  \item K-Nearest Neighbors (\textbf{KNN})~\cite{k_nearest_neighbor} predicts  by taking the majority vote among the $k$ nearest instances
  \item Logistic Regression (\textbf{LR})~\cite{lr} predicts class probabilities using a weighted linear combination of features
  \item Multilayer Perceptron (\textbf{MLP})~\cite{mlp} predicts using a feedforward neural network with one or more hidden layers 
  \item Random Forest (\textbf{RF})~\cite{random_forest} predicts through majority voting based on an ensemble of decision trees combined via bagging
  \item Support Vector Machine (\textbf{SVM})~\cite{svc} identifies a maximum-margin separating hyperplane to distinguish instances, via linear or nonlinear kernels.
\end{itemize}
\looseness=-1

The models are trained to predict a comprehensibility metric or proxy (\eg PBU) given a code snippet $c$ inspected by programmer $p$, as represented by a vector of (concatenated) features: syntactic and developer features in the \textit{\textbf{Control}} setting, and the same features supplemented with the semantic feature in \textit{\textbf{Treatment}} setting. We do this for each model, proxy, and corresponding dataset.

\subsection{Data Preprocessing}
\label{subsec:data_normalization_balancing}

We applied three standard preprocessing steps: duplicate removal, normalization, and class balancing.

First, duplicates were removed from the training data to prevent overfitting.  Duplicate instances in the validation/test sets were preserved to account for realistic model usage.
Second, feature values were standardized using $z$-score normalization~\cite{Abdi:ERD10}, transforming values $x$ into $z=\frac{x-\mu}{\sigma}$ to ensure comparability across scales.  
Finally, to address class imbalance (see \cref{tab:absolute_proxies}), we applied Synthetic Minority Over-sampling (SMOTE)~\cite{chawla2002smote} to generate synthetic samples for minority classes in the training data only.

Because the number of features exceeded the number of data instances, we applied correlation-based feature selection (as in~\cite{Scalabrino:TSE19}) before oversampling. Features were ranked by Kendall’s $\tau$~\cite{kendall1938new} correlation with a target comprehensibility metric, and we evaluated models using the top 10\%, 20\%, \ldots, 100\% of features. 
Due to their small number, all developer features were retained.

In the treatment setting, we augmented the previously selected features with the verifier warning sum.

\subsection{Model Training and Evaluation}
\label{subsec:model_training_evaluation}

To reduce overfitting and bias in relatively small datasets, we used nested cross-validation (CV)~\cite{cross_validation}, which tunes and evaluates models iteratively across folds, providing better generalization~\cite{Cawley:JMLR2010}.

Nested CV consists of two levels: the \textbf{outer} 10-fold CV splits data into train/test sets, maintaining class balance. Each fold serves once as the test set. The \textbf{inner} 5-fold CV optimizes model hyperparameters within each outer training set. 
We tested multiple hyperparameter combinations (see the replication package~\cite{repl_pack}), starting from values used in prior work~\cite{Scalabrino:TSE19} and refining them based on established guidelines~\cite{Breiman:Springer2001,Amy:Medium22}. 
Optimal parameters were selected by the highest weighted F1-score (see \cref{subsec:metrics}). We selected models with multiple optimal configurations across folds as this minimizes bias compared to selecting a single best set, as done in prior work~\cite{Scalabrino:TSE19}.

To balance accuracy and training time, we chose 10 outer folds
and 5 inner folds. More folds reduce test set size, making results less
reliable, while fewer folds risk overfitting. After determining the
best hyperparameter sets across all 10 outer training sets, we trained
the models using each optimal set and evaluated them on every test
fold. Unlike prior work~\cite{Scalabrino:TSE19}, which selects a single best hyperparameter set, our approach reduces bias by testing multiple optimal
configurations, leading to a more reliable performance estimates.
\looseness=-1

\input{latex_tables/compare_table}

\subsection{Evaluation Metrics}
\label{subsec:metrics}

Model performance was evaluated using standard classification metrics. We began by computing \textit{precision (P)}, \textit{recall (R)}, and \textit{F1-score (F1)}~\cite{hossin:IJDKP2015,Naidu:Springer23}.  
Scores were computed per class ($P_i$, $R_i$, $F1_i$) and aggregated into weighted averages ($wP$, $wR$, $wF1$) based on the class distribution of each comprehensibility proxy.
We computed global metrics by summing confusion matrix entries across folds before calculation~\cite{Forman:SigKDD2010}, avoiding distortions from across-fold averaging.
Additionally we used the Area Under the Receiver Operating Characteristic Curve (AUC-ROC)~\cite{hanley1982meaning}, which measures a model's ability to classify instances across all classification thresholds, providing a comprehensive measure of predictive power. As with the other metrics, AUC-ROC is computed per classes and aggregated into weighted AUC-ROC ($wAUC$), based on the class distribution of each proxy. For space reasons, we only report summary metrics ($wF1$ and $wAUC$) in \cref{tab:code_comprehensibility_results_comparision,tab:code_comprehensibility_results_comparision_per_tool}, but the results based on all the metrics are found in our replication package~\cite{repl_pack}.

Since we compare the absolute metric differences between control and treatment settings, across the six model types, five comprehensibility metrics, and different sets of model hyperparameters, we employed the non-parametric, unpaired Mann-Whitney U test~\cite{mann-witney} to assess statistical significance (evaluated at a confidence level of 95\%). Given a particular Metric ($wF1$ or $wAUC$), the null hypothesis ($H_{0}$) posits that $C_{Metric} \geq T_{Metric}$ and, hence, the alternative hypothesis ($H_{a}$) posits that $T_{Metric} > C_{Metric}$. $C_{Metric}$ and $T_{Metric}$ represent model 
performance obtained from the control and treatment settings, respectively.

%% file: latex_tables/code_features.tex
\begin{table*}[t]
  \centering
  \caption{Syntactic features categorized by type.}
  \label{tab:code_features}
  \resizebox{0.85\textwidth}{!}{%
  \begin{tabular}{l|l|c}
    \toprule
  \textbf{Category} & \textbf{Examples}                                    & \multicolumn{1}{l}{\textbf{\# Features}} \\ \hline
  Complexity        & Cyclomatic complexity, \# of nested blocks, \# of loops, \# of comparisons, entropy, \ldots    & 10 \\
  Size              & LOC, \# of parameters, \# of statements, \# of literals, \# of assignments, \# of numbers, \ldots   & 17 \\
  Lexicon           & \# of identifiers, \# of keywords, Identifier length, \# of operators, text coherence, \ldots & 27 \\
  Format            & \# of blank lines, \# of spaces, \# of parentheses, \# of commas, \# of periods, line length, \ldots  & 18 \\
  Documentation     & \# of comments, comment readability, comments and identifier consistency, \ldots & 12 \\ \hline
  \multicolumn{2}{c|}{\textbf{Total}} & \textbf{84} \\                                           
    \bottomrule
\end{tabular}
  }
\end{table*}

%% file: latex_tables/compare_table.tex
\begin{table*}[t]
\caption{Code comprehensibility prediction results based on wF1 and wAUC, averaged across trained models. \Code (Control setting): Syntactic + Developer features; \CW (Treatment setting): Syntactic + Developer + Semantic features; $\Delta$ = \CW – \Code; A positive difference $\Delta$ is highlighted in \hlgreen{Green}. Statistically significant differences ($p < 0.05$) are in bold.}
\label{tab:code_comprehensibility_results_comparision}
\begin{subtable}[c]{\textwidth}
   \centering
   \caption{$wF1$ results for each metric and model}
    \resizebox{\textwidth}{!}{%
        \begin{tabular}{l|rrc|rrc|rrc|rrc|rrc|rrc} 
        \toprule
        \multirow{2}{*}{\textbf{Metric}} & \multicolumn{3}{c|}{\textbf{NB}}                                                                         & \multicolumn{3}{c|}{\textbf{KNN}}                                                                        & \multicolumn{3}{c|}{\textbf{LR}}                                                                         & \multicolumn{3}{c|}{\textbf{MLP}}                                                                        & \multicolumn{3}{c|}{\textbf{RF}}                                                                         & \multicolumn{3}{c}{\textbf{SVM}}                                                                         \\
                                        & \multicolumn{1}{c}{\Code} & \multicolumn{1}{c}{\CW} & \multicolumn{1}{c|}{\DIFF} & \multicolumn{1}{c}{\Code} & \multicolumn{1}{c}{\CW} & \multicolumn{1}{c|}{\DIFF} & \multicolumn{1}{c}{\Code} & \multicolumn{1}{c}{\CW} & \multicolumn{1}{c|}{\DIFF} & \multicolumn{1}{c}{\Code} & \multicolumn{1}{c}{\CW} & \multicolumn{1}{c|}{\DIFF} & \multicolumn{1}{c}{\Code} & \multicolumn{1}{c}{\CW} & \multicolumn{1}{c|}{\DIFF} & \multicolumn{1}{c}{\Code} & \multicolumn{1}{c}{\CW} & \multicolumn{1}{c}{\DIFF}  \\ 
        \hline
        \AU               & 0.273               & 0.273                 & 0.000                                & 0.278                              & 0.278                           & 0.000                              & 0.324                          & 0.319                           & -0.004                             & 0.255                         & 0.227                           & -0.028                            & 0.340                          & 0.344                          & \cellcolor{lightgreen}0.004         & 0.304                             & 0.296                            & -0.008                        \\
        \PBU              & 0.556               & 0.555                 & -0.001                               & 0.537                              & 0.524                           & -0.013                             & 0.593                          & 0.595                           & \cellcolor{lightgreen}0.002        & 0.516                         & 0.510                           & -0.005                            & 0.666                          & 0.674                          & \cellcolor{lightgreen}0.008         & 0.584                             & 0.588                            & \cellcolor{lightgreen}0.004   \\
        \ABUFIF           & 0.600               & 0.599                 & -0.001                               & 0.559                              & 0.553                           & -0.006                             & 0.614                          & 0.611                           & -0.003                             & 0.489                         & 0.460                           & -0.030                            & 0.611                          & 0.605                          & -0.006                              & 0.589                             & 0.580                            & -0.009                         \\
        \BDFIF            & 0.579               & 0.598                 & \cellcolor{lightgreen}\textbf{0.018}           & 0.567                             & 0.542                           & -0.025                             & 0.572                          & 0.564                           & -0.008                             & 0.573                         & 0.614                           & \cellcolor{lightgreen}0.042       & 0.702                          & 0.699                          & -0.003                              & 0.565                             & 0.557                            & -0.008                         \\ 
        \hline
        \RL               & 0.179               & 0.187                 & \cellcolor{lightgreen}\textbf{0.007}         & 0.199                               & 0.194                           & -0.005                             & 0.178                          & 0.183                           & \cellcolor{lightgreen}0.005        & 0.202                         & 0.201                           & \textbf{-0.001}                            & 0.162                          & 0.168                          & \cellcolor{lightgreen}\textbf{0.007}         & 0.165                             & 0.169                            & \cellcolor{lightgreen}0.004    \\
        \bottomrule
        \end{tabular}
    } 
\end{subtable}
\begin{subtable}[c]{\textwidth}
    \label{tab:code_comprehensibility_results_comparision_auc}
    \centering
    \vspace{0.2cm}
    \caption{$wAUC$ results for each metric and model}
    \resizebox{\textwidth}{!}{%
    	\begin{tabular}{l|rrc|rrc|rrc|rrc|rrc|rrc}
        \toprule
        \multirow{2}{*}{\textbf{Metric}} & \multicolumn{3}{c|}{\textbf{NB}}                                                                         & \multicolumn{3}{c|}{\textbf{KNN}}                                                                        & \multicolumn{3}{c|}{\textbf{LR}}                                                                         & \multicolumn{3}{c|}{\textbf{MLP}}                                                                        & \multicolumn{3}{c|}{\textbf{RF}}                                                                         & \multicolumn{3}{c}{\textbf{SVM}}                                                                         \\
                                        & \multicolumn{1}{c}{\Code} & \multicolumn{1}{c}{\CW} & \multicolumn{1}{c|}{\DIFFAUC} & \multicolumn{1}{c}{\Code} & \multicolumn{1}{c}{\CW} & \multicolumn{1}{c|}{\DIFFAUC} & \multicolumn{1}{c}{\Code} & \multicolumn{1}{c}{\CW} & \multicolumn{1}{c|}{\DIFFAUC} & \multicolumn{1}{c}{\Code} & \multicolumn{1}{c}{\CW} & \multicolumn{1}{c|}{\DIFFAUC} & \multicolumn{1}{c}{\Code} & \multicolumn{1}{c}{\CW} & \multicolumn{1}{c|}{\DIFFAUC} & \multicolumn{1}{c}{\Code} & \multicolumn{1}{c}{\CW} & \multicolumn{1}{c}{\DIFFAUC}  \\ 
        \hline
        \AU               & 0.552                & 0.548                & -0.004                & 0.535                & 0.534                & -0.001                 & 0.575                & 0.572                & -0.003                & 0.514                & 0.503                & -0.011                & 0.586                & 0.583                & -0.003                & 0.560                & 0.552                & -0.009                \\
        \PBU              & 0.511                & 0.506                & -0.005                & 0.528                & 0.526                & -0.002                & 0.528                & 0.529                & \cellcolor{lightgreen}0.001                 & 0.502                & 0.500                & -0.002                & 0.556                & 0.558                & \cellcolor{lightgreen}0.001                 & 0.526                & 0.528                & \cellcolor{lightgreen}0.001                 \\
        \ABUFIF           & 0.497                & 0.498                & \cellcolor{lightgreen}0.001                 & 0.498                & 0.498                & 0.000                 & 0.497                & 0.497                & 0.000                 & 0.499                & 0.500                & \cellcolor{lightgreen}\textbf{0.001}                 & 0.496                & 0.496                & 0.000                 & 0.498                & 0.498                & 0.000                 \\
        \BDFIF            & 0.460                & 0.455                & -0.005                & 0.462                & 0.462                & 0.000                 & 0.471                & 0.477                & \cellcolor{lightgreen}\textbf{0.006}                 & 0.512                & 0.484                & -0.028                & 0.437                & 0.438                & \cellcolor{lightgreen}0.001                 & 0.490                & 0.495                & \cellcolor{lightgreen}0.006                 \\ 
        \hline
        \RL               & 0.490                & 0.494                & \cellcolor{lightgreen}\textbf{0.004}                 & 0.522                & 0.518                & -0.003                & 0.503                & 0.512                & \cellcolor{lightgreen}\textbf{0.009}                 & 0.508                & 0.519                & \cellcolor{lightgreen}\textbf{0.011}                 & 0.489                & 0.492                & \cellcolor{lightgreen}\textbf{0.003}                 & 0.507                & 0.514                & \cellcolor{lightgreen}0.007                 \\
        \bottomrule
        \end{tabular}
    }
\end{subtable} 
\end{table*}

%% file: 4_results.tex
\section{Results and Analysis}
\label{sec:code_comprehensibility_results}

Given the combinatorial nature of our experimental design, we ended up training 1,228 classifiers in the control setting (\ie models trained solely on syntactic and developer features) and 935 in the treatment setting (\ie models trained on the augmented semantic feature). These total encompass models trained across six machine learning types, ten feature sets, five comprehensibility metrics, and multiple hyperparameter combinations optimized via nested cross-validation (see \cref{subsec:model_training_evaluation}).

The \Code (``\textbf{\emph{Control}}'') columns in \cref{tab:code_comprehensibility_results_comparision} report the aggregated performance of models trained solely on syntactic and developer features, and \CW (``\textbf{\emph{Treatment}}'') columns display the results when using the semantic feature alongside syntactic and developer features. Results are grouped by metric and model type. Model performance is expressed as the average weighted F1-score ($wF1$) and weighted AUC ($wAUC$) across all trained classifiers. We focus our analysis of $wF1$ results as we obtained similar trends with the $wAUC$ metric.
\looseness=-1

\subsection{Analysis across Metrics and Model Types}

We first discuss the results across comprehensibility metrics and models types for both control and treatment settings (see \cref{tab:code_comprehensibility_results_comparision}).

Among the five comprehensibility metrics, \PBU (Perceived Binary Understandability), \ABUFIF (Actual Binary Understandability) and \BDFIF (Binary Deceptiveness) are the easiest to predict in both control and treatment settings, consistently achieving the highest scores across all model types. When using only the syntactic and developer features,  model $wF1$ scores range from 0.489 to 0.702, and supplementing these features with the semantic feature leads to 0.46 to 0.699 $wF1$ scores.

Notably, \AU (Actual Understandability) and \RL (Readability Level) are more difficult to predict than the other three metrics.  Between these two, \RL is the most challenging metric to predict regardless of features, obtaining $wF1$ scores ranging from 0.162 to 0.202 in control models, and 0.168 to 0.201 in treatment models. This is confirmed by $wAUC$ results, which shows model performance lower for \RL than for \AU. We attribute this to two main factors. (1) \textit{Multi-Class Classification Challenge:} Both metrics involve multiple predictive classes (4 for \AU and 5 for \RL), which are inherently more difficult to predict than binary metrics~\cite{Del:Access2022}. (2) \textit{Subjectivity and Noise:}  \RL relies on subjective ratings provided directly by developers, introducing greater variability and noise compared to objective measures like \AU. The decision boundaries between adjacent ratings (\eg 2 vs. 3) might be subtle, leading to inconsistencies even in human judgement.

Overall, Random Forests (RF) achieved the highest scores across most metrics in both control and treatment settings, with a few exceptions. In the control setting using only syntactic and developer features, RF was slightly below the top-performing models: about 0.5\% lower than LR on \ABUFIF and 25\% lower than MLP on \RL. A similar pattern appears when augmenting verifier warnings, where RF trailed LR by roughly 1\% on \ABUFIF and achieved 0.168 on \RL approximately 20\% lower than the best-performing MLP.
As an ensemble model combining multiple decision trees, RF is capable of capturing complex, non-linear patterns in the data, making it better-suited for the diverse nature of code comprehensibility metrics.

As shown in \cref{tab:code_comprehensibility_results_comparision_auc}, both control and treatment models demonstrate some capacity to learn from the data, as their AUC scores outperform the na\"ive random baseline ($wAUC=0.5$). At the same time, 47\% of models fail to effectively learn meaningful patterns and underperform even the naïve baseline, showing mixed results across models and metrics.
Additionally, both $wF1$ and $wAUC$ scores remain generally low across models and evaluation metrics, suggesting that these models provide limited reliability as predictors of code comprehensibility. Our results are in line with Scalabrino \etal~\cite{Scalabrino:TSE19}, who concluded that these ML models are far from being practical.
\looseness=-1

\subsection{Analysis of Control vs Treatment Models} 

\Cref{tab:code_comprehensibility_results_comparision} shows that incorporating the aggregated count of verifier warnings as a semantic feature does not yield any meaningful improvement. 
Across the 30 model–metric combinations, only 10-13 exhibit any gain depending on the evaluation metric, and even those are marginal and mostly not statistically significant ($\Delta wF1 = +0.002$ to $+0.042$, and $\Delta wAUC = +0.001$ to $+0.009$). 
56.6\% to 66.7\% of all configurations show either no benefit or an outright decline when verifier warnings are added, although with most of the degradation being not statistically significant either.

The largest improvement, observed for \BDFIF with the MLP model, reaches only $\Delta wF1 = +0.042$, a small 7.2\% relative increase over the control models (trained solely with syntactic and developer features). 
Although every metric except \ABUFIF shows a slight improvement in at least one isolated configuration, no metric exhibits consistent gains across models.
\RL is the only metric that shows improvement in four out of six models, although the gains are small ($\Delta wF1$ ranging from +0.004 to +0.007).
These results are consistent with the correlation findings reported by Feldman \etal~\cite{Feldman:FSE23}, which showed that the \textit{verifier warning sum} exhibits a small correlation with \RL, whereas \PBU and \AU show no correlation.
While using features that correlate with the target metric is not the only factor enabling models to learn underlying patterns, it plays a significant role. Here, the correlation of \RL with the \textit{verifier warning sum} may explain the consistent improvements in \RL observed across models when this feature is added.

Overall, these findings suggest that while verifier-warning counts may occasionally capture patterns related to code comprehensibility when models happen to be well-calibrated, their predictive contribution is generally weak and largely negligible. As such, 
\textbf{adding \textit{verifier warning sum} does not improve ML models’ ability to predict
	code comprehensibility} in a meaningful way.
	\looseness=-1

\input{latex_tables/tool_wise_results}

\subsubsection{Analysis across model types and metrics.}

We analyze the results across model types and metrics to understand this overall finding. RF, SVM, LR and NB remain the most robust performers, achieving $wF1$ and $wAUC$ values up to 0.699 and 0.583 (for treatment), respectively.
Having improvements in two to three metrics in these models indicate that adding verifier warnings provides improvements, but marginally. 
The inclusion of warnings yields limited but occasionally positive effects, suggesting that while they may capture certain comprehensibility patterns, their overall predictive contribution remains weak.

The metrics sensitive to user perception (\PBU and \RL), show small but consistent improvements across three or four models (RF, SVM, LR, and NB) when verifier warnings are included. 
This suggests that metrics driven by user ratings and developers' perceived understanding may benefit more from semantic information about the code than purely objective metrics. For example, when code contains issues that verifiers detect, such as pointer dereferences without null checks or array accesses that are not obviously in-bounds, developers may perceive these as more severe, which may influence their judgments of code comprehensibility. Such information is not fully captured by syntactic features alone.
Furthermore, the \RL metric originates from the \dsthree dataset, which contains simplified code snippets designed to reduce complexity. In this setting, syntactic features alone may be insufficient to capture the subtle factors influencing developers' ratings in these simplified contexts. The addition of semantic features like verifier warnings may help bridge this gap.

In order to understand why incorporating \textit{verifier warning sum} as an additional semantic feature does not improve model performance, we conducted a statistical analysis between this feature and each comprehensibility metric.
Rather than relying on a single statistical indicator, we examined the feature from three complementary perspectives: (i) rank-based association using Kendall's $\tau$~\cite{Kendall1938}, (ii) general statistical dependency using Mutual Information (MI)~\cite{shannon1948mathematical, cover1999elements}, and (iii) model-level contribution using SHAP (SHapley Additive exPlanations) values~\cite{lundberg2017unified}. 

Kendall's rank correlation tests whether a monotonic relationship exists between the feature and each comprehensibility metric. A near-zero Kendall's $\tau$ coefficient indicates the absence of a consistent directional association, irrespective of the underlying data distribution.
Mutual Information captures statistical dependencies that extend beyond monotonic relationships, including non-linear and threshold-based effects. 
A low MI score indicates that the feature and a metric are largely statistically independent, providing no meaningful discriminative information.
SHAP value analysis was conducted as a post-hoc model explanation technique to evaluate how the trained classifiers utilized the feature at the individual prediction level. Unlike the preceding statistical measures, SHAP examines the learned model directly, attributing a feature contribution value for each data instance.

All three analyses yielded consistently weak results. Kendall's $\tau$ coefficients were really small (between -0.19 and 0.03), indicating no monotonic association of the \textit{verifier warning sum} feature with each metric. MI scores were similarly low, suggesting statistical near-independence between the feature and the metrics. SHAP value analysis further confirmed this, with the feature exhibiting near-zero SHAP magnitudes, inconsistent directionality across data instances, and unstable importance rankings across classes and cross-validation folds. For \AU classes 0 and 2, the SHAP values are approximately zero, and \textit{verifier warning sum} is ranked last among the top 10\% features (i.e., twelfth place). In contrast, for classes 1 and 3, the SHAP values are slightly higher but still remain close to zero; \textit{verifier warning sum} is ranked fifth among the same 12 features, with SHAP values ranging between -0.02 and 0.02. This suggests each class instance in the metric is influenced differently inside the model when other features are present.
The convergence of these analyses provides evidence that \textit{verifier warning sum} carries no reliable or learnable signal for comprehensibility prediction, and its inclusion does not contribute meaningfully to model performance.
Please refer to our replication package~\cite{repl_pack} for detailed results of these analyses, including the full set of Kendall's $\tau$ coefficients, MI scores, and SHAP value distributions for all features across all models and metrics.

\subsection{Verifier-Specific Warning Analysis}
\label{sec:tool_specific_analysis}

To examine whether modeling verifier-specific semantic information yields better predictive performance than aggregating warnings across all tools, we conducted the same control–treatment comparison using the warning count of each verifier. In other words, rather than combining warnings from all four verifiers into a single aggregated feature, we treated each tool’s warning count as an individual semantic feature to assess whether tool-specific signals improve prediction accuracy.

We focus our analysis in this subsection on the $wF1$ metric. 
According to the results presented in  \cref{tab:code_comprehensibility_results_comparision_per_tool}, among the four tools, OpenJML (OJ) and Java Typestate Checker  (JaTyC) exhibit at least one $wF1$ improvement across all six models and all five metrics. Specifically, for each metric, the treatment values obtained using OJ and JaTyC warnings outperform the control setting in at least one of the evaluated models, with $\Delta wF1$ ranging from $+0.001$ to $+0.05$.
\looseness=-1

Similarly, \PBU, \BDFIF, and \RL are the comprehensibility metrics that benefit most from adding individual verifier warnings, showing $wF1$ improvements in at least two models whenever any individual verifier warning count is used as a feature (with one exception: when use warning count from Infer, PBU improves in only one treatment model), with $\Delta wF1$ scores ranging from $+0.001$ to $+0.05$.

When examining the predictive power of individual verifiers, JaTyC and the Checker Framework (CF) yield the highest number of statistically significant improvements in $\Delta wF1$, whereas the Infer verifier yields none. Within these statistically significant combinations, the \RL metric emerges as the most robust, driving improvements across multiple models (most notably NB, LR, and SVM) for both the CF and JaTyC verifiers. While these significant gains for the \RL metric are largely mirrored in the $\Delta AUC$ results, $\Delta AUC$ also exposes a distinct trend absent from the $\Delta wF1$ data: a minor but statistically significant improvement for the \ABUFIF metric when paired with the MLP model, which occurs consistently across all four verifiers.

However, no single verifier consistently outperforms the others across all metrics and models. This suggests that the predictive value of warnings depends on the specific categories of issues each tool detects in the code and how those issues relate to different dimensions of code comprehensibility.
It is worth mentioning that, using individual verifier warnings from CF, OJ, and JaTyC leads to improvements across almost all models, although the gains are minimal. This is explained by the fact that CF, OJ, and JaTyC issue comparatively more warnings (CF issues 51 and 70 warnings for \dsthree and \dssix, respectively. The corresponding numbers for OJ are 605 and 236, and for JaTyC, 327 and 522), suggesting they capture a broader range of semantic information, whereas Infer issues fewer warnings (0 for \dsthree and only 5 for \dssix), which may explain its smaller improvements.

When using individual verifier warning counts as semantic features, RF and SVM remain the most robust performers, achieving $\Delta wF1$ values of up to 0.022. 
RF leads to improvements across all the verifiers for \AU and \RL (except for Infer), although the improvements are minimal ($\Delta wF1$ $0.001$ to $0.014$). SVM shows improvements for \PBU and \RL across all verifiers, with $\Delta wF1$ values ranging from $+0.005$ to $+0.022$.

Overall, as observed with the \textit{verifier warning sum}, incorporating individual verifier warnings produces limited but occasionally positive effects. In most cases, the improvements are modest, typically not exceeding $\Delta wF1 = +0.05$ (when using the MLP model with OpenJML and Checker Framework warnings to predict \BDFIF). 
These results confirm that the earlier conclusion derived from the aggregated warning sum also holds when individual verifier warnings are used as separate semantic features.

%% file: latex_tables/tool_wise_results.tex
\begin{table*}[t]
    \centering
    \caption{Code comprehensibility prediction results per verifier. Model performance differences are shown: $\Delta$ = \CW – \Code;  \Code (Control setting): Syntactic + Developer features; \CW (Treatment setting): Syntactic + Developer + Semantic features; The semantic feature is the warning count of each individual verifier. A positive difference $\Delta$ is highlighted in \hlgreen{Green}. Statistically significant differences ($p < 0.05$) are in bold.}
    \label{tab:code_comprehensibility_results_comparision_per_tool}
    \begin{subtable}{0.48\textwidth}
        \centering
        \caption{$\Delta wF1$ results for each verifier, metric, and model}
        \label{tab:f1_score_tool_wise}
        \resizebox{\columnwidth}{!}{%
            \begin{tabular}{l|l|c|c|c|c|c|c}
                \toprule
                \multicolumn{1}{c|}{\textbf{Verifier}} & \multicolumn{1}{c|}{\textbf{Metric}} & \multicolumn{1}{c|}{\textbf{NB}} & \multicolumn{1}{c|}{\textbf{KNN}} & \multicolumn{1}{c|}{\textbf{LR}} & \multicolumn{1}{c|}{\textbf{MLP}} & \multicolumn{1}{c|}{\textbf{RF}} & \multicolumn{1}{c}{\textbf{SVM}}  \\
                \hline
                \multirow{5}{*}{Infer}   & \AU                         & \cellcolor{lightgreen}0.006    & -0.001                  & -0.002                              & -0.029                        & \cellcolor{lightgreen}0.007       & -0.003                   \\
                                        & \PBU                        & 0.000                          & -0.003                  & 0.000                               & -0.022                        & 0.000                             & \cellcolor{lightgreen}0.005                    \\
                                        & \ABUFIF                     & -0.009                         & -0.009                  & -0.001                              & -0.030                        & -0.003                            & -0.017                   \\
                                        & \BDFIF                      & \cellcolor{lightgreen}0.001    & -0.009                  & 0.000                               & \cellcolor{lightgreen}0.038   & \cellcolor{lightgreen}0.001       & \cellcolor{lightgreen}0.010                    \\
                                        & \RL                         & 0.000                          & -0.003                  & 0.000                               & -0.011                        & 0.000                             & \cellcolor{lightgreen}0.014                    \\
                \hline
                \multirow{5}{*}{CF}      & \AU                         & -0.005                         & -0.006                  & -0.003                              & -0.028                        & \cellcolor{lightgreen}0.007       & -0.004                   \\
                                        & \PBU                        & -0.002                         & -0.003                  & \cellcolor{lightgreen}0.001         & -0.022                        & \cellcolor{lightgreen}0.001       & \cellcolor{lightgreen}\textbf{0.010}                    \\
                                        & \ABUFIF                     & -0.003                         & -0.008                  & -0.002                              & -0.028                        & -0.005                            & -0.001                   \\
                                        & \BDFIF                      & 0.000                          & -0.011                  & -0.004                              & \cellcolor{lightgreen}0.050   & -0.003                            & \cellcolor{lightgreen}\textbf{0.011}                    \\
                                        & \RL                         & \cellcolor{lightgreen}\textbf{0.006}    & -0.004                  & \cellcolor{lightgreen}\textbf{0.008}         & \cellcolor{lightgreen}\textbf{0.002}   & \cellcolor{lightgreen}\textbf{0.005}       & \cellcolor{lightgreen}\textbf{0.022}                    \\
                \hline
                \multirow{5}{*}{OJ}      & \AU                         & -0.004                         & -0.006                  & -0.003                              & -0.027                        & \cellcolor{lightgreen}0.003       & \cellcolor{lightgreen}0.002                    \\
                                        & \PBU                        & \cellcolor{lightgreen}0.007    & -0.024                  & \cellcolor{lightgreen}0.003         & -0.009                        & \cellcolor{lightgreen}\textbf{0.008}       & \cellcolor{lightgreen}0.007                    \\
                                        & \ABUFIF                     & \cellcolor{lightgreen}0.001    & -0.005                  & \cellcolor{lightgreen}0.002         & -0.026                        & -0.005                            & -0.003                   \\
                                        & \BDFIF                      & -0.005                         & -0.012                  & -0.005                              & \cellcolor{lightgreen}0.050   & -0.006                            & \cellcolor{lightgreen}0.003                    \\
                                        & \RL                         & \cellcolor{lightgreen}0.002    & 0.000                   & \cellcolor{lightgreen}0.004         & -0.002                        & \cellcolor{lightgreen}0.004       & \cellcolor{lightgreen}\textbf{0.015}                    \\
                \hline                         
                \multirow{5}{*}{JaTyC}   & \AU                         & 0.000                          & 0.000                   & -0.002                              & -0.029                        & \cellcolor{lightgreen}\textbf{0.011}       & -0.001                   \\
                                        & \PBU                        & 0.000                          & -0.006                  & \cellcolor{lightgreen}0.002         & -0.002                        & -0.005                            & \cellcolor{lightgreen}\textbf{0.014}                    \\
                                        & \ABUFIF                     & \cellcolor{lightgreen}0.003    & -0.007                  & -0.003                              & -0.028                        & 0.000                             & -0.001                   \\
                                        & \BDFIF                      & \cellcolor{lightgreen}\textbf{0.012}    & -0.020                  & -0.014                              & \cellcolor{lightgreen}0.040   & -0.008                            & -0.001                   \\
                                        & \RL                         & \cellcolor{lightgreen}\textbf{0.008}    & -0.002                  & \cellcolor{lightgreen}\textbf{0.006}         & \cellcolor{lightgreen}0.002   & \cellcolor{lightgreen}\textbf{0.014}       & \cellcolor{lightgreen}\textbf{0.019}                    \\
                \bottomrule
            \end{tabular}
        } 
    \end{subtable}
    \hfill
    \begin{subtable}{0.48\textwidth}
        \centering
        \caption{$\Delta wAUC$ results for each verifier, metric, and model}
        \label{tab:auc_score_tool_wise}
        \resizebox{\columnwidth}{!}{%
            \begin{tabular}{l|l|c|c|c|c|c|c}
            \toprule
            \multicolumn{1}{c|}{\textbf{Verifier}} & \multicolumn{1}{c|}{\textbf{Metric}} & \multicolumn{1}{c|}{\textbf{NB}} & \multicolumn{1}{c|}{\textbf{KNN}} & \multicolumn{1}{c|}{\textbf{LR}} & \multicolumn{1}{c|}{\textbf{MLP}} & \multicolumn{1}{c|}{\textbf{RF}} & \multicolumn{1}{c}{\textbf{SVM}}  \\
            \hline
            \multirow{5}{*}{Infer}       & \AU                         & -0.006                         & -0.004                        & -0.002                        & -0.010                        & -0.001                        & -0.010                   \\
                                        & \PBU                        & 0.000                           & 0.000                         & -0.001                        & -0.005                        & 0.000                         & -0.002                   \\
                                        & \ABUFIF                     & \textbf{0.000}                           & 0.000                         & 0.000                         & \cellcolor{lightgreen}\textbf{0.001}   & 0.000                         & 0.000                    \\
                                        & \BDFIF                      & \cellcolor{lightgreen}\textbf{0.005}     & \cellcolor{lightgreen}0.002   & \cellcolor{lightgreen}0.004   & -0.024                        & \cellcolor{lightgreen}0.002   & \cellcolor{lightgreen}0.006                    \\
                                        & \RL                         & 0.000                           & -0.001                        & 0.000                         & \cellcolor{lightgreen}0.003   & -0.001                        & -0.001                   \\
            \hline
            \multirow{5}{*}{CF}          & \AU                         & -0.006                         & -0.010                        & 0.000                         & -0.010                        & 0.000                         & -0.008                   \\
                                        & \PBU                        & -0.005                          & -0.006                        & -0.001                        & -0.005                        & -0.002                        & \cellcolor{lightgreen}0.001                    \\
                                        & \ABUFIF                     & 0.000                           & 0.000                         & 0.000                         & \cellcolor{lightgreen}\textbf{0.001}   & \textbf{0.000}                         & 0.000                    \\
                                        & \BDFIF                      & \cellcolor{lightgreen}0.002     & \cellcolor{lightgreen}0.003   & \cellcolor{lightgreen}0.004   & -0.028                        & \cellcolor{lightgreen}\textbf{0.003}   & \cellcolor{lightgreen}0.004                    \\
                                        & \RL                         & \cellcolor{lightgreen}\textbf{0.012}     & \cellcolor{lightgreen}0.002   & \cellcolor{lightgreen}\textbf{0.021}   & \cellcolor{lightgreen}\textbf{0.014}   & \cellcolor{lightgreen}\textbf{0.013}                         & \cellcolor{lightgreen}\textbf{0.018}                    \\
            \hline
            \multirow{5}{*}{OJ}          & \AU                         & -0.005                         & -0.005                        & -0.002                        & -0.009                        & -0.004                        & -0.006                   \\
                                        & \PBU                        & -0.001                          & -0.002                        & 0.000                         & -0.003                        & \cellcolor{lightgreen}0.001   & 0.000                    \\
                                        & \ABUFIF                     & 0.000                           & 0.000                         & 0.000                         & \cellcolor{lightgreen}\textbf{0.001}   & 0.000                         & 0.000                    \\
                                        & \BDFIF                      & \cellcolor{lightgreen}0.004     & -0.006                        & \cellcolor{lightgreen}0.005   & -0.029                        & \cellcolor{lightgreen}0.002   & \cellcolor{lightgreen}0.006                    \\
                                        & \RL                         & 0.000                           & -0.003                        & 0.000                         & \cellcolor{lightgreen}0.003   & -0.001                        & -0.001                   \\
            \hline
            \multirow{5}{*}{JaTyC}       & \AU                         & -0.002                         & \cellcolor{lightgreen}0.001   & -0.004                        & -0.010                        & \cellcolor{lightgreen}0.002   & -0.006                   \\
                                        & \PBU                        & -0.005                          & \cellcolor{lightgreen}0.002   & 0.000                         & -0.004                        & -0.001                        & \cellcolor{lightgreen}0.002                    \\
                                        & \ABUFIF                     & 0.000                           & 0.000                         & 0.000                         & \cellcolor{lightgreen}\textbf{0.001}   & 0.000                         & 0.000                    \\
                                        & \BDFIF                      & -0.007                          & -0.003                        & \cellcolor{lightgreen}\textbf{0.008}   & -0.029                        & \cellcolor{lightgreen}\textbf{0.004}   & \cellcolor{lightgreen}\textbf{0.011}                    \\
                                        & \RL                         & \cellcolor{lightgreen}\textbf{0.004}     & -0.001                        & \cellcolor{lightgreen}\textbf{0.007}   & \cellcolor{lightgreen}\textbf{0.012}                         & \cellcolor{lightgreen}\textbf{0.017}   & \cellcolor{lightgreen}0.005                    \\
            \bottomrule
            \end{tabular}
        } 
    \end{subtable}
    \vspace{0.1cm} %
    \footnotesize %
    \\
    Note: For improved readability, individual results for both Control and Treatment settings are omitted but can be found in our replication package~\cite{repl_pack}.
\end{table*}

%% file: 8_threats_to_validity.tex
\section{Threats to Validity}
\label{sec:threats}
\textbf{Construct Validity.} We rely on human-judged comprehensibility metrics (\eg perceived understandability and readability) as proxies for the complex, multifaceted cognitive process of code comprehension. 
It is possible that these proxies do not fully capture all dimensions of code comprehensibility, such as the influence of domain knowledge or individual differences in cognitive styles.
To minimise this, threat we used a wider range of comprehensibility proxies, including both subjective and objective measures, to capture different aspects of code comprehensibility.

We use the counts of verifier warnings as a proxy for the actual information processed by the verifiers. 
It is possible that warning counts do not perfectly encapsulate the semantic nuances that affect code complexity and human cognitive load. However, to ensure methodological robustness, we employed the same proxy metric successfully used and validated in prior work~\cite{Feldman:FSE23}.

\textbf{Internal Validity.} The choice of models and their hyperparameters, our feature selection procedure, and our cross-validation methodology are validity threats.
We also built our dataset of code features (\cref{subsec:ac_data_sources}) from the definitions of Scalabrino \etal~\cite{Scalabrino:TSE19}. 
Some definitions were ambiguous, and we found discrepancies between the data generated by their code and the data provided in their replication package. We discarded a few code features and implemented the rest using the Java 8 Language Specification~\cite{javaspec} to resolve ambiguities.

Although the datasets we used in this study are the largest available Java comprehensibility datasets used in prior work, the limited number of instances could impact the power of our statistical analyses and the ability of the models to learn patterns from the data. However, the $wAUC$ results show that the models are indeed learning from the data, although not consistently across models and comprehensibility metrics.
\looseness=-1

\textbf{External Validity.} All the snippets from \dssix and \dsthree datasets are written in Java. The results may not generalize to code written in other programming languages. To our best knowledge, comprehensibility datasets for other languages are scarce, or contain made-up code not representative of real-world code. In contrast, the selected datasets are extracted from OSS projects, thus representing real-life code.
The majority of the study participants from whom both comprehensibility  data were obtained in prior work are mostly students. Only \dsthree included some professional developers. The results may not generalize to broader populations of professional developers. Further, the results may not generalize to larger code snippets.
\looseness=-1

%% file: 6_conclusions.tex
\section{Implications, Conclusions, and Future work}
\label{sec:conclusions}

Our results showed mostly no performance improvement of verifier-augmented ML models compared to baseline models that solely rely on syntactic and developer features, with only a small handful of cases exhibiting marginal and inconsistent gains.

These negative findings highlight \textbf{two important insights}:

\begin{enumerate}
\item Developers construct a rich cognitive model when understanding code, one that is not fully captured by the syntactic, semantic, and developer-specific features used in this study, which are also commonly used in the current literature~\cite{Scalabrino:TSE19,Posnett:MSR2011,Raymond:TSE10}. This suggests that additional, yet-unidentified factors play a substantial role in shaping human perceptions of code comprehensibility.

\item Combining syntactic and developer features is sufficient for predicting code comprehensibility at the state-of-the-art accuracy level we obtained, which is consistent with the accuracy reported in prior studies~\cite{Scalabrino:TSE19}. 
This could mean that their combination may be enough to capture code complexity, that \emph{none} of the features are useful to predict comprehensibility, or that the semantic feature we studied (verifier warning sum) does not fully capture nuanced aspects of code semantics and complexity.
\looseness=-1
\end{enumerate}

Future work holds several promising avenues for researchers. 
Efforts should focus on conducting more comprehensive user studies to uncover new influential factors that could be used as model features. %
As warning counts do not improve mode performance significantly, this feature may not be an appropriate way to represent code complexity. Our future work will experiment with alternative representations, such as the types of warnings issued by the verifiers or the types of facts about the code that verifiers prove during code analysis, which can provide a more nuanced view of code semantics and complexity. For example, richer semantic representations from program analysis (\eg control-flow or data-flow relations) may yield superior results compared to  warning counts. 
For practitioners,  our findings suggest that existing machine learning models are not yet reliable enough for automated code comprehensibility estimation in real-world development. Before such tools can be confidently applied, further methodological advances and program semantic modeling will be necessary.

%% file: 7_acknowledgements.tex
\section*{Acknowledgments}

This work is supported in part by NSF grants 2414110, 2414111, and 2239107. Any opinions, findings, and conclusions expressed herein are the authors' and do not necessarily reflect those of the sponsors.
\looseness=-1